\documentclass{article}

\usepackage{PRIMEarxiv}

\usepackage[utf8]{inputenc} 
\usepackage[T1]{fontenc}    
\usepackage{hyperref}       
\usepackage{url}            
\usepackage{booktabs}       
\usepackage{amsfonts}       
\usepackage{nicefrac}       
\usepackage{microtype}      
\usepackage{lipsum}
\usepackage{fancyhdr}       
\usepackage{graphicx}       
\graphicspath{{media/}}     
\usepackage[most]{tcolorbox}
\usepackage{enumitem}
\usepackage{multirow}
\usepackage{pdflscape}
\usepackage{longtable}
\usepackage{cleveref}

\title{A Comprehensive Study on the Use of Word Embedding Models in Software Engineering Domain
}

\author{
  Xiaohan Chen, Weiqin Zou, Lianyi Zhi, Qianshuang Meng, Jingxuan Zhang \\
  Nanjing University of Aeronautics and Astronautics \\
  Nanjing {\rm 210016}, China\\
  \texttt{\{xiaohanchen, weiqin, lianyi.zhi, qs\_meng, jxzhang\}@nuaa.edu.cn} \\
}

\begin{document}
\maketitle

\begin{abstract}
Word embedding (WE) techniques are advanced textual semantic representation models oriented from the natural language processing (NLP) area. Inspired by their effectiveness in facilitating various NLP tasks, more and more researchers attempt to adopt these WE models for their software engineering (SE) tasks, of which semantic representation of software artifacts such as bug reports and code snippets is the basis for further model building. However, existing studies are generally isolated from each other without comprehensive comparison and discussion. This not only makes the best practice of such cross-discipline technique adoption buried in scattered papers, but also makes us kind of blind to current progress in the semantic representation of SE artifacts. To this end, we decided to perform a comprehensive study on the use of WE models in the SE domain. 181 primary studies published in mainstream software engineering venues are collected for analysis. Several research questions related to the SE applications, the training strategy of WE models, the comparison with traditional semantic representation methods, etc., are answered. With the answers, we get a systematical view of the current practice of using WE for the SE domain, and figure out the challenges and actions in adopting or developing practical semantic representation approaches for the SE artifacts used in a series of SE tasks.
\end{abstract}


\section{Introduction}
\label{intro_sect}
Word embedding (WE) is an advancement in the natural language processing (NLP) area that makes computers better understand text-based content. As a type of word representation, it is considered one remarkable breakthrough of deep learning in solving challenging NLP problems\cite{1}. With WE models, words are represented in real-valued numeric vectors. These vectors embed individual words into a feature space (hence the name word embeddings) with generally a few hundred dimensions; they are expected to capture the context of a word in a document, its semantic and syntactic features, its relation with other words, etc.\cite{2}. Compared to other traditional word representations, such as the bag of words, WE vectors are relatively low-dimensional. Moreover, as the vectors are learned based on word usage, this makes words with similar meanings have similar vector values and, hence, naturally become close-by in the n-dimensional geometric feature space\cite{3}.

As a revolutionary word-meaning-capturing tool through low-dimensional numeric vectors, WEs have demonstrated great potential in facilitating various NLP tasks based on text analysis, such as text classification, named entity recognition, question answering, machine translation, etc.\cite{4,5,6}. Both the academic and industrial communities have been and are still devoting much effort to developing a series of more advanced WE models, such as ELMo\cite{7} and BERT\cite{8}. Meanwhile, it is also becoming more and more prevalent for practitioners from other disciplines attempting to leverage the WE achievements in the NLP area to handle their tasks at hand. A representative cross-discipline usage is to apply WE models for the software engineering (SE) domain.

As an important subcategory of computer science, one major goal of the software engineering discipline is to develop various SE techniques to help practitioners better develop and maintain software products. Correspondingly, the demand for well-performed SE techniques to ensure software quality is increasing in the software-defined age\cite{9}. These SE techniques generally rely on the analysis of different kinds of software artifacts generated in the software development and maintenance activities, including source code, documentation, bug reports, specifications, etc. All these artifacts need to be digitized to make computers understand. Whether the semantics embedded in the artifacts could be well represented largely affects the performance of those SE techniques.

In the early stage, traditional information retrieval models such as the vector space model (VSM), latent semantic analysis (LSA), latent dirichlet allocation (LDA), abstract syntactic tree (AST), etc., are used to extract semantics from software artifacts. These models generally consider little about the contextual semantics of tokens/words/terms or may fail to extract hidden high-level semantics within the artifacts. Inspired by the potential of deep learning and WE in the NLP area, researchers propose to adapt WE models (originally used to represent the semantics of plain texts) to process software artifacts. Taking the WE vectors of software artifacts as data basis, by applying machine learning or other model-building algorithms, a series of automatic SE techniques are built, such as bug localization, test case generation, API recommendations, etc.\cite{10,11,12,13}.

However, existing studies that used WE models for the SE domain are generally isolated from each other without comprehensive comparison and discussion. This makes the best practice of such cross-discipline technique adoption buried in scattered papers. Further, it also keeps us from obtaining a general view of current progress in the semantic representation of SE artifacts. Considering the key role of semantic representation of SE artifacts, we decide to perform a systematical analysis of the use of WE models for the SE domain. Specifically, we first retrieved 1,957 candidate studies through designed search keywords upon 45 software engineering venues. Then we obtained 181 primary studies for analysis after applying our inclusion and exclusion criteria. We designed four research questions that relate to different aspects of the practice of using WE models for the SE domain, including the involved SE applications/artifacts, the training strategy of WE models, the comparison with traditional semantic representation methods, etc. Through the analysis, we found that: (1) The adoption of word embedding models in the field of software engineering has been on a rising trend year by year (except for the year of 2023); (2) Software maintenance and development are two main areas where WE models are applied in relevant SE tasks; (3) Word2Vec and BERT are the top two models used in the SE domain, with SE-specific embeddings being generally favored over generic models; and (4) there generally lacks a systematic comparative analysis in the selection of WE models. Through our study, we get a comprehensive understanding of the current practice of using WE for the SE domain, and provide some actional suggestions in adopting or developing practical semantic representation approaches for the SE artifacts used in a series of SE tasks.

The remaining parts of our paper are structured as follows. Section~\ref{method_sect} introduces the review methods we adapted to perform this study. Section~\ref{res_sect} presents the results. The discussion and related work are described in Section~\ref{discuss_sect} and ~\ref{related_sect}. Finally, we conclude our study in Section~\ref{conclude_sect}.

\section{Methodology}
\label{method_sect}
In this section, we first introduce the process of searching and identifying relevant papers, including the search keywords and the inclusion and exclusion criteria we used to search papers. Then we describe the research questions we aim to answer to understand the use of WE models in the SE domain.
\subsection{Search Strategy}
\subsubsection{Selected Venues}
As our goal is to understand the use of WE models in the SE domain, we mainly select the papers from the venues that appear in the list of international academic periodicals and conferences recommended by the China Computer Federation (CCF), more specifically belong to the Software Engineering/Systems Software/Programming Language category in the list. Moreover, to make it actionable for manual checking with individual paper content, we mainly take into account the venues ranked as A or B in the list (venues with C-rank are excluded). In total, 45 venues are considered while searching relevant papers, including 29 conferences such as ICSE and FSE, and 16 journals such as TSE and TOSEM. The detailed venues and retrieved relevant papers can be referred in \href{https://docs.qq.com/sheet/DZFZtQ053TVlpT2tH?tab=pe37sc}{https://docs.qq.com/sheet/WE in SE}.
\subsubsection{Keywords}
Since our focus is on the WE models for the SE domain, the first keyword we naturally used is ``word embedding''. At first, We used ``word embedding'' to search within the selected venues on the DBLP website (\href{http://dblp.uni-trier.de/}{http://dblp.uni-trier.de/}). Then, we randomly selected several retrieved papers and checked their contents. We found that some papers would directly use the names of WE models (such as Word2Vec, BERT, etc.) in the whole paper rather than using ``word embedding''. In addition to the keyword ``word embedding'', we also add the names of existing WE models as search keywords, namely ``word2vec'', ``GloVe'', ``fasttext'', ``BERT'' and ``ELMo''. In other words, for each selected venue, we would search its official website through DBLP with the above six keywords separately, to retrieve any paper whose full-text contains at least one keyword. In this step, we obtain 1,957 candidate papers. The time for literature search is up to 2023, including early access articles.
\subsubsection{Inclusion/Exclusion Criteria}
After collecting the initial 1,957 papers, two authors manually checked each paper to identify whether it was a relevant one according to the following inclusion criteria and exclusive criteria.
\begin{itemize}
    \item The paper aims to propose an automatic SE technique, of which semantic representation is a key step and the representation approach used is a specific WE model.
    \item Papers that do semantic representation beyond word embedding granularity but at, for example, sentence level or document level, are excluded.
    \item The paper is a regular research paper. Workshop/Symposium/industry/demonstration papers, posters, etc., are excluded.
    \item Review studies, e.g., literature review or survey, are excluded.
    \item If a conference paper is extended to a journal article, we only keep the journal version.
\end{itemize}
\subsection{Research Questions}
In this study, we aim to, on one hand, obtain an overall view of the adoption of WE models for semantic representation in various SE tasks and, on the other hand, identify potential research opportunities for adopting and developing practical semantic representation techniques for SE domain. To achieve these objectives, we design the following four research questions whose answers may help us understand different aspects of the use of WE models in the SE domain.

\textbf{RQ1. What is the distribution of the studies across publication years and venues?} (To obtain the trend of the publications in the domain).

\textbf{RQ2. What SE tasks tend to use WE models for semantic representation?} (To reveal the prevalence of WE use in different SE tasks, hence identify potential application opportunities).

\textbf{RQ 3. What WE models are generally adopted by SE tasks, and are they compared with other semantic representation models in the evaluation experiments?} (To understand the strengths and limitations of these models in real-world SE applications, and potentially revealing gaps in evaluation practices or model selection that could improve SE task performance.)

\textbf{RQ4. What is the general way to obtain WE vectors in SE tasks? By using the general pre-trained WE models or training a domain-specific one?} (To provide insights into which training approach better captures SE-specific semantics and improves task outcomes.)

\section{Results}
\label{res_sect}
\subsection{RQ1. What is the distribution of the studies across publication years and venues?}

\begin{figure}
    \centering
    \includegraphics[width=0.8\linewidth]{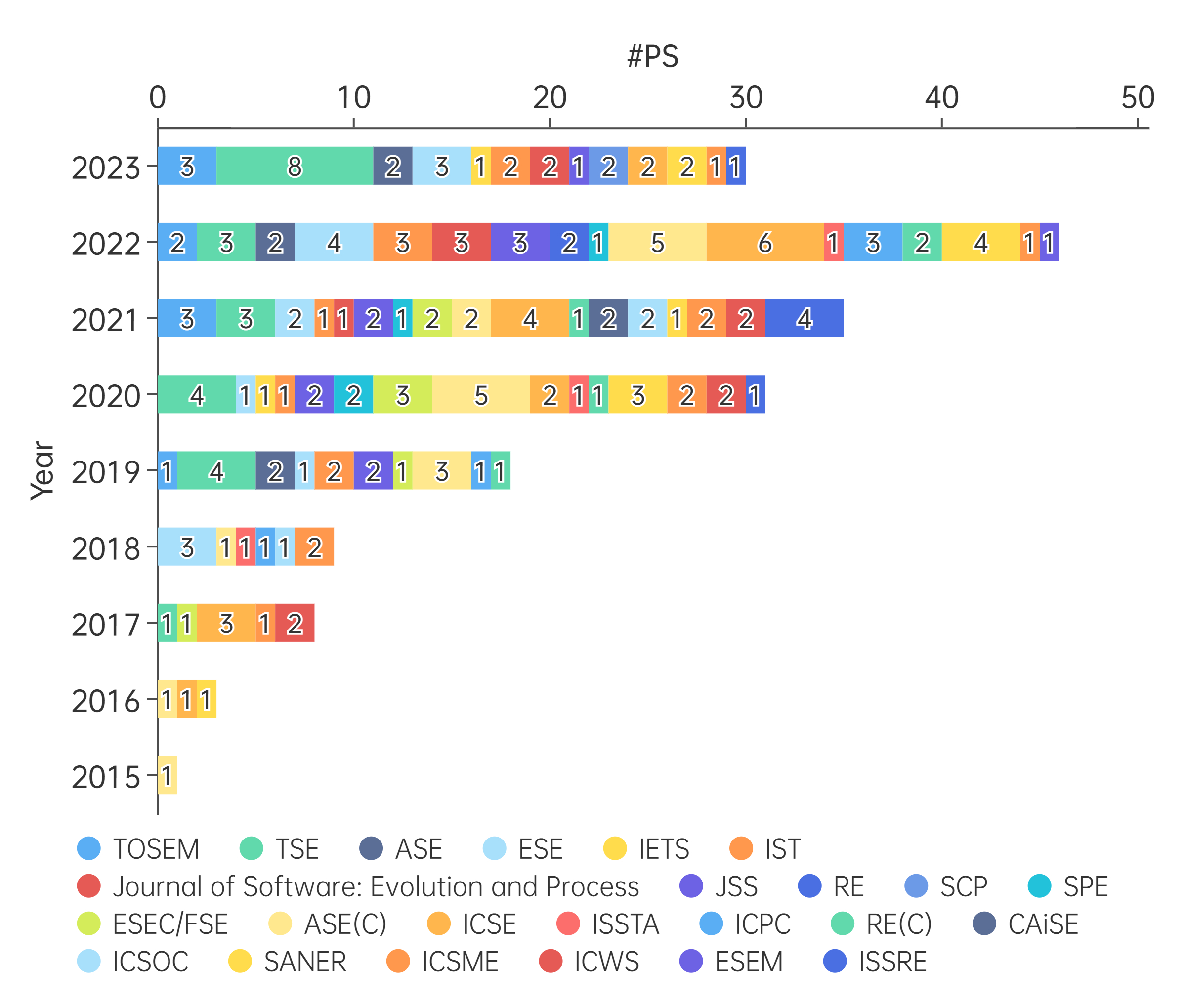}
    \caption{Primary Studies (\#PS) by year and publication.}
    \label{fig:distribution}
\end{figure}

To answer RQ1, after obtaining the 181 primary studies (PS) from various sources, we collected their publication years and the journals or conferences where they were published. Figure~\ref{fig:distribution} shows the distribution of these studies across different years and journals/conferences.

From the publication year view, we can find that the first paper that applied WE model in the SE domain is published in the year of 2015 (Phong et al.\cite{14} applied Word2Vec to the opinion mining task in the SE domain. Word2Vec\cite{15} is the first WE model released by Google in 2013). Between 2015 and 2017, only the Word2Vec model was applied to tasks in the SE domain. In 2018, researchers began to use other WE models (e.g., GloVe\cite{16} (released in 2014), fastText\cite{17} (released in 2016)) to solve SE-related tasks. Since then, up until 2022, the number of papers using WE techniques in the SE domain has grown significantly, though there was a slight decline in 2023. This may be attributed to the increasing popularity of Seq2Seq models (with embedding layers integrated into the model) and the emergence of large language models such as GPT. However, this does not imply a decrease in the application of WE techniques; rather, it has, to some extent, expanded the concept of traditional word embeddings, suggesting that context-aware embedding methods may become increasingly prevalent. These show that the application of WE technology in the SE domain is currently an active research area.

From the venues view, we can find that among all publications, \textit{IEEE Transactions on Software Engineering (TSE, CCF-Rank A)} published the largest number of related studies, with a total of 23. \textit{International Conference on Automated Software Engineering (ASE(C), CCF-Rank A)} and \textit{International Conference on Software Engineering (ICSE ,CCF-Rank A)} ranked second in terms of contribution, both publishing 18, followed by \textit{Empirical Software Engineering (ESE, CCF-Rank B)} with 14 studies. These figures indicate that the cross-disciplinary use of word embedding models in the software engineering domain is gaining increasing recognition within the SE research community.

\begin{tcolorbox}
    [colback=gray!10,
    colframe=black!80,
    arc=2mm, auto outer arc,
    title={RQ1 - Summary},breakable,
    before upper={\parindent15pt\noindent},]
    
    The first use of WE model in the SE domain was in 2015. Few studies were conducted from 2016 to 2018. During the period 2019 to 2023, the application of WE in SE has expanded to considerable research interest. Meanwhile, TSE, ICSE and ASE are the top three venues published the most papers related to WE adoption in SE domain.
\end{tcolorbox}

\subsection{RQ2. What SE tasks tend to use WE models for semantic representation?}
In this RQ, we attempt to achieve two sub goals, one is to obtain a general view of the concrete SE tasks that adopted WE models for semantic representation by providing a taxonomy of those SE tasks. The other one is to understand exactly what kinds of software artifacts tend to more likely be represented by WE models for semantic representation.

\subsubsection{Taxonomy of SE Tasks}

We followed the open coding practice\cite{18} by manually applying codes (i.e., SE tasks—e.g., API recommendation, vulnerability detection, test automation) to the studies (in a shared online spreadsheet). We first checked the keywords and abstracts of each paper. If a certain keyword explained the SE task solved by the paper, we selected it as the extracted code. This process was carried out by two authors independently. If the code of an article was uncertain, the two authors would read the introduction or even the whole paper till the code was determined through discussion. Next, the authors discussed together conceptually-related codes by generalizing or specializing them, employing the Qualitative Content Analysis approach. After these processes, the accuracy of the determination of SE tasks was guaranteed.

After identifying the SE task addressed by each paper, we combined both closed and open card sorting methods to develop a taxonomy of those SE tasks. First, for the closed card sorting part, we pre-identified four categorical themes based on the Software Life Cycle (SLC), i.e., requirement engineering, software development, software testing, and software maintenance, and made cards labeled with SE tasks. Then, two authors worked independently to place these cards into the four pre-defined theme categories. In some cases, two authors may place a paper in different pre-defined categories, in this situation, they would discuss together to determine its final category. It happens that some papers may not be covered by the pre-defined theme categories. In this case, the authors would conduct open card sorting to assign new theme categories to them. Specifically, they would independently create new theme codes for these papers, and then discuss together to determine their categories. 

The final results are shown in \Cref{tab:tasks_1,tab:tasks_2,tab:tasks_3,tab:tasks_4} that present the taxonomy of 181 papers (The 181 references that started with prefix `R' in the tables, and the complete taxonomy introductions on SE tasks could be found at the website: \href{https://docs.qq.com/doc/DZE5YbmpiR0d0VUJr}{\textit{Taxonomy of SE Tasks}}.). The number of papers covered by the four pre-defined categories in the SLC is 142 (78\% of all studies). For the remaining SE tasks, the two authors established two new category themes (i.e., general task support and project management) through Open Card Sorting, covering 25 and 14 primary studies, respectively. Among the six areas, the software maintenance area has the highest number of tasks utilizing WE technology for semantic representation of software artifacts, accounting for 66 papers (36\%). This is followed by the software development area, which has 44 papers (24\%). Both areas are critical stages in the software lifecycle, involving numerous related software artifacts such as code and bug reports. In the \textit{``Software Maintenance''} area, the most common subarea, ``Defect Handling,'' accounts for 42 tasks (23\%), encompassing defect detection, localization, fixing, and severity analysis. 
Tasks such as ``Bug/Vulnerability Detection'' and ``Bug/Fault Localization'' are the most frequently represented in terms of semantic representation using WE within detection and localization subcategories.
In the subarea of ``Code Quality Evaluation \& Optimization,'' tasks like ``Code Review'' and ``SATD Detection'' are particularly notable. The main subareas utilizing WE for semantic representation in \textit{``Software Development''} include ``Code Entity Recommendation/Generation'' and ``Code Comprehension.'' In ``Code Entity Recommendation/Generation,'' the API category is predominant, featuring tasks like ``API Recommendation'' (8 studies), ``API Mapping'' (3 studies), and ``Similar Technology Comparison'' (1 study). Other notable subcategories include ``Code Examples,'' with tasks such as ``Code Examples Recommendation'' (3 studies), and related categories like ``Log.'' The ``Code Comprehension'' subarea includes tasks such as ``API Knowledge Retrieval'' (5 studies), ``Code Summarization'' (2 studies), and ``API Extraction \& Linking'' (3 studies). These tasks play a crucial role in enhancing developers' understanding and management of code during the software development process. Additionally, there is a certain percentage of studies addressing the use of WE technology for semantic representation in SE tasks related to requirements engineering and software testing, accounting for 10\% and 7\% of the primary studies, respectively. The results suggest that within \textit{``Requirements Engineering,''} WE technology is predominantly used for tasks related to requirement acquisition and management. While for \textit{``Software Testing,''} it emphasizes tasks that improve automation and testing reliability, reflecting a growing research interest in leveraging WE to enhance the efficiency and consistency of software testing processes.

\begin{landscape}
    \aboverulesep=0pt
    \belowrulesep=0pt
    \begin{longtable}{@{}p{20mm}|p{12mm}|p{28mm}|p{38mm}|p{78mm}|p{35mm}@{}}
    \caption{The Taxonomy of SE tasks that Adopt WE models for semantic representation (1/4)}
    \label{tab:tasks_1}\\
    \toprule[1.5pt]
    \textbf{SLC area} & \textbf{PS} & \textbf{Subarea} & \multicolumn{2}{l|}{\textbf{SE task}} & \textbf{Reference} \\* \midrule[1.5pt]
    \endhead
    \multirow{11}{*}{\begin{tabular}[c]{@{}l@{}}Requirement \\ Engineering\end{tabular}} & \multirow{11}{*}{19(10\%)} & \multirow{2}{*}{\begin{tabular}[c]{@{}l@{}}Requirement \\ Elicitation\end{tabular}} & \multicolumn{2}{l|}{Requirement Acquisition} & R{[}87, 127, 171, 174{]} \\* \cmidrule(l){4-6} 
     &  &  & \multicolumn{2}{l|}{App Review Analysis} & R{[}20, 104{]} \\* \cmidrule(l){3-6} 
     &  & \multirow{3}{*}{\begin{tabular}[c]{@{}l@{}}Requirement \\ Modelling\end{tabular}} & \multicolumn{2}{l|}{Text Annotation} & R{[}144{]} \\* \cmidrule(l){4-6} 
     &  &  & \multicolumn{2}{l|}{Domain Model Autocomplete} & R{[}145{]} \\* \cmidrule(l){4-6} 
     &  &  & \multicolumn{2}{l|}{GUI Prototyping} & R{[}38{]} \\* \cmidrule(l){3-6} 
     &  & \multirow{3}{*}{\begin{tabular}[c]{@{}l@{}}Requirement \\ Management\end{tabular}} & \multicolumn{2}{l|}{Requirements Classification} & R{[}37, 64, 100, 139{]} \\* \cmidrule(l){4-6} 
     &  &  & \multicolumn{2}{l|}{Requirement Linking} & R{[}77{]} \\* \cmidrule(l){4-6} 
     &  &  & \multicolumn{2}{l|}{Feature Redundancy Detection} & R{[}135{]} \\* \cmidrule(l){3-6} 
     &  & \multirow{3}{*}{\begin{tabular}[c]{@{}l@{}}Requirement \\ Verification\end{tabular}} & \multicolumn{2}{l|}{Semantic Web Generation} & R{[}81{]} \\* \cmidrule(l){4-6} 
     &  &  & \multicolumn{2}{l|}{Entity Coreference Detection} & R{[}80, 140{]} \\* \cmidrule(l){4-6} 
     &  &  & \multicolumn{2}{l|}{Cross-Domain Ambiguity Detection} & R{[}36{]} \\* \midrule
    \multirow{14}{*}{\begin{tabular}[c]{@{}l@{}}Software \\ Development\end{tabular}} & \multirow{14}{*}{44(24\%)} & \multirow{14}{*}{\begin{tabular}[c]{@{}l@{}}Code Entity \\ Recommendation/\\ Generation\end{tabular}} & \multicolumn{1}{l|}{\multirow{5}{*}{API}} & API Recommendation & \begin{tabular}[c]{@{}l@{}}R{[}2, 22, 66, 83, 96, \\ 146, 158, 165{]}\end{tabular} \\* \cmidrule(l){5-6} 
     &  &  & \multicolumn{1}{l|}{} & API Comparison & R{[}105{]} \\* \cmidrule(l){5-6} 
     &  &  & \multicolumn{1}{l|}{} & API Mapping & R{[}11, 89, 124{]} \\* \cmidrule(l){5-6} 
     &  &  & \multicolumn{1}{l|}{} & Analogical Libraries Recommendation & R{[}41, 154{]} \\* \cmidrule(l){5-6} 
     &  &  & \multicolumn{1}{l|}{} & Similar Technology Comparison \& Differencing & R{[}21{]} \\* \cmidrule(l){4-6} 
     &  &  & \multicolumn{1}{l|}{\multirow{2}{*}{Code Examples}} & Differentiated Code Retrieval \& Recommendation & R{[}51, 134{]} \\* \cmidrule(l){5-6} 
     &  &  & \multicolumn{1}{l|}{} & Code Examples Recommendation & R{[}23, 70, 162{]} \\* \cmidrule(l){4-6} 
     &  &  & \multicolumn{1}{l|}{Task Solution} & Task Solution Recommendation & R{[}47{]} \\* \cmidrule(l){4-6} 
     &  &  & \multicolumn{1}{l|}{\multirow{3}{*}{Log}} & Logging Variables Recommendation & R{[}14{]} \\* \cmidrule(l){5-6} 
     &  &  & \multicolumn{1}{l|}{} & Logging Location Recommendation & R{[}102{]} \\* \cmidrule(l){5-6} 
     &  &  & \multicolumn{1}{l|}{} & Log Level Recommendation & R{[}86{]} \\* \cmidrule(l){4-6} 
     &  &  & \multicolumn{2}{l|}{Type Inference} & R{[}92, 111, 121, 152{]} \\* \cmidrule(l){4-6} 
     &  &  & \multicolumn{2}{l|}{App Permission Recommendation} & R{[}35, 172{]} \\* \cmidrule(l){4-6} 
     &  &  & \multicolumn{2}{l|}{Program Generation} & R{[}138{]} \\* \bottomrule[1.5pt]
    \end{longtable}
\end{landscape}
\begin{landscape}
    \aboverulesep=0pt
    \belowrulesep=0pt
    \begin{longtable}{@{}p{20mm}|p{12mm}|p{28mm}|p{38mm}|p{68mm}|p{35mm}@{}}
    \caption{The Taxonomy of SE tasks that Adopt WE models for semantic representation (2/4)}
    \label{tab:tasks_2}\\
    \toprule[1.5pt]
    \textbf{SLC area} & \textbf{PS} & \textbf{Subarea} & \multicolumn{2}{l|}{\textbf{SE task}} & \textbf{Reference} \\* \midrule[1.5pt]
    \endhead
    \multirow{6}{*}{\begin{tabular}[c]{@{}l@{}}Software \\ Development\end{tabular}} & \multirow{6}{*}{44(24\%)} & \multirow{6}{*}{\begin{tabular}[c]{@{}l@{}}Code \\ Comprehension\end{tabular}} & \multicolumn{2}{l|}{Code Summarization} & R{[}30, 156{]} \\* \cmidrule(l){4-6} 
     &  &  & \multicolumn{2}{l|}{Program Comprehension} & R{[}32{]} \\* \cmidrule(l){4-6} 
     &  &  & \multicolumn{2}{l|}{Code Comment Scope Detection} & R{[}75{]} \\* \cmidrule(l){4-6} 
     &  &  & \multicolumn{2}{l|}{API Extraction \& Linking} & R{[}13, 45, 119{]} \\* \cmidrule(l){4-6} 
     &  &  & \multicolumn{2}{l|}{API Type Resolution} & R{[}82{]} \\* \cmidrule(l){4-6} 
     &  &  & \multicolumn{2}{l|}{API Knowledge Retrieval} & \begin{tabular}[c]{@{}l@{}}R{[}8, 48, 137, 155, \\ 173{]}\end{tabular} \\* \midrule
    \multirow{8}{*}{\begin{tabular}[c]{@{}l@{}}Software \\ Testing\end{tabular}} & \multirow{8}{*}{13(7\%)} & \multirow{4}{*}{Test Automation} & \multicolumn{1}{l|}{\multirow{3}{*}{Test Cases Generation}} & Test Inputs Generation & R{[}103, 123, 166{]} \\* \cmidrule(l){5-6} 
     &  &  & \multicolumn{1}{l|}{} & Test Oracle Generation & R{[}128, 132{]} \\* \cmidrule(l){5-6} 
     &  &  & \multicolumn{1}{l|}{} & Test Cases Design & R{[}109{]} \\* \cmidrule(l){4-6} 
     &  &  & \multicolumn{2}{l|}{Test Transfer} & R{[}98, 106{]} \\* \cmidrule(l){3-6} 
     &  & \multirow{3}{*}{Test Maintenance} & \multicolumn{2}{l|}{Test Cases Maintenance} & R{[}31, 153{]} \\* \cmidrule(l){4-6} 
     &  &  & \multicolumn{2}{l|}{Flaky Test Cases Prediction} & R{[}24{]} \\* \cmidrule(l){4-6} 
     &  &  & \multicolumn{2}{l|}{Test Report Consistency Detection} & R{[}29{]} \\* \cmidrule(l){3-6} 
     &  & \begin{tabular}[c]{@{}l@{}}Test Models \\ Building\end{tabular} & \multicolumn{2}{l|}{Test Models Building} & R{[}59{]} \\* \midrule
    \multirow{7}{*}{\begin{tabular}[c]{@{}l@{}}Software \\ Maintenance\end{tabular}} & \multirow{7}{*}{66(36\%)} & \multirow{7}{*}{Defects Handling} & \multicolumn{1}{l|}{\multirow{7}{*}{Detection}} & Bug/Vulnerability Detection & \begin{tabular}[c]{@{}l@{}}R{[}17, 28, 61, 68, 78, \\ 88, 122, 133{]}\end{tabular} \\* \cmidrule(l){5-6} 
     &  &  & \multicolumn{1}{l|}{} & Log Anomaly Detection & R{[}99, 118, 176, 181{]} \\* \cmidrule(l){5-6} 
     &  &  & \multicolumn{1}{l|}{} & API-related Compatibility Issues Detection & R{[}125{]} \\* \cmidrule(l){5-6} 
     &  &  & \multicolumn{1}{l|}{} & Privacy Compliance Violations Detection & R{[}110{]} \\* \cmidrule(l){5-6} 
     &  &  & \multicolumn{1}{l|}{} & Conflict Detection & R{[}170{]} \\* \cmidrule(l){5-6} 
     &  &  & \multicolumn{1}{l|}{} & Bug Prediction & R{[}62{]} \\* \cmidrule(l){5-6} 
     &  &  & \multicolumn{1}{l|}{} & Log Analytics & R{[}129, 130{]} \\* \bottomrule[1.5pt]
    \end{longtable}
\end{landscape}
\begin{landscape}
    \aboverulesep=0pt
    \belowrulesep=0pt
    \begin{longtable}{@{}p{20mm}|p{12mm}|p{28mm}|p{38mm}|p{78mm}|p{35mm}@{}}
    \caption{The Taxonomy of SE tasks that Adopt WE models for semantic representation (3/4)}
    \label{tab:tasks_3}\\
    \toprule[1.5pt]
    \textbf{SLC area} & \textbf{PS} & \textbf{Subarea} & \multicolumn{2}{l|}{\textbf{SE task}} & \textbf{Reference} \\* \midrule[1.5pt]
    \endhead
    \multirow{24}{*}{\begin{tabular}[c]{@{}l@{}}Software \\ Maintenance\end{tabular}} & \multirow{24}{*}{66(36\%)} & \multirow{15}{*}{Defects Handling} & \multicolumn{1}{l|}{\multirow{7}{*}{Localization}} & Bug/Fault Localization & \begin{tabular}[c]{@{}l@{}}R{[}46, 49, 57, 120, \\ 178{]}\end{tabular} \\* \cmidrule(l){5-6} 
     &  &  & \multicolumn{1}{l|}{} & Bug Report Enhancement/Reformulation & R{[}79, 142, 159{]} \\* \cmidrule(l){5-6} 
     &  &  & \multicolumn{1}{l|}{} & Issue Report Classification & R{[}55{]} \\* \cmidrule(l){5-6} 
     &  &  & \multicolumn{1}{l|}{} & Security Bug Report Detection & R{[}177{]} \\* \cmidrule(l){5-6} 
     &  &  & \multicolumn{1}{l|}{} & Bug Report Entity Recognition & R{[}34, 73{]} \\* \cmidrule(l){5-6} 
     &  &  & \multicolumn{1}{l|}{} & Linked Incident Identification & R{[}91{]} \\* \cmidrule(l){5-6} 
     &  &  & \multicolumn{1}{l|}{} & Bug Reproduction & R{[}3{]} \\* \cmidrule(l){4-6} 
     &  &  & \multicolumn{1}{l|}{\multirow{5}{*}{Fix}} & Automated Program Repair & R{[}26, 112{]} \\* \cmidrule(l){5-6} 
     &  &  & \multicolumn{1}{l|}{} & Vulnerability Fix Detection & R{[}108{]} \\* \cmidrule(l){5-6} 
     &  &  & \multicolumn{1}{l|}{} & Bug Fixing Time Prediction & R{[}53{]} \\* \cmidrule(l){5-6} 
     &  &  & \multicolumn{1}{l|}{} & Patch Identification & R{[}5, 157{]} \\* \cmidrule(l){5-6} 
     &  &  & \multicolumn{1}{l|}{} & Bug Triage & R{[}90{]} \\* \cmidrule(l){4-6} 
     &  &  & \multicolumn{1}{l|}{\multirow{3}{*}{\begin{tabular}[c]{@{}l@{}}Severity \\ Analysis\end{tabular}}} & Severity Level Prediction of CVEs & R{[}7{]} \\* \cmidrule(l){5-6} 
     &  &  & \multicolumn{1}{l|}{} & Vulnerability Severity Prediction & R{[}160{]} \\* \cmidrule(l){5-6} 
     &  &  & \multicolumn{1}{l|}{} & Issue Priority Prediction & R{[}175{]} \\* \cmidrule(l){3-6} 
     &  & \multirow{9}{*}{\begin{tabular}[c]{@{}l@{}}Code Quality \\ Evaluation \& \\ Optimization\end{tabular}} & \multicolumn{1}{l|}{\multirow{4}{*}{\begin{tabular}[c]{@{}l@{}}Code \\ Quality \\ Evaluation\end{tabular}}} & Code Review & R{[}9, 16, 150, 168{]} \\* \cmidrule(l){5-6} 
     &  &  & \multicolumn{1}{l|}{} & Code Clone Detection & R{[}107, 131, 136{]} \\* \cmidrule(l){5-6} 
     &  &  & \multicolumn{1}{l|}{} & SATD Detection & R{[}1, 42, 50, 95{]} \\* \cmidrule(l){5-6} 
     &  &  & \multicolumn{1}{l|}{} & Code Readability Classification & R{[}72{]} \\* \cmidrule(l){4-6} 
     &  &  & \multicolumn{1}{l|}{\multirow{5}{*}{\begin{tabular}[c]{@{}l@{}}Code \\ Optimization\end{tabular}}} & Identifier Normalization & R{[}27{]} \\* \cmidrule(l){5-6} 
     &  &  & \multicolumn{1}{l|}{} & SATD Removal Strategies Recommendation & R{[}149{]} \\* \cmidrule(l){5-6} 
     &  &  & \multicolumn{1}{l|}{} & Software Design Pattern Detection & R{[}40, 74{]} \\* \cmidrule(l){5-6} 
     &  &  & \multicolumn{1}{l|}{} & Binaries Optimization & R{[}25{]} \\* \cmidrule(l){5-6} 
     &  &  & \multicolumn{1}{l|}{} & Linguistic Anti-patterns Detection & R{[}39{]} \\* \bottomrule[1.5pt]
     \end{longtable}
\end{landscape}
\begin{landscape}
    \aboverulesep=0pt
    \belowrulesep=0pt
    \begin{longtable}{@{}p{20mm}|p{12mm}|p{28mm}|p{38mm}|p{78mm}|p{35mm}@{}}
    \caption{The Taxonomy of SE tasks that Adopt WE models for semantic representation (4/4)}
    \label{tab:tasks_4}\\
    \toprule[1.5pt]
    \textbf{SLC area} & \textbf{PS} & \textbf{Subarea} & \multicolumn{2}{l|}{\textbf{SE task}} & \textbf{Reference} \\* \midrule[1.5pt]
    \endhead
    \multirow{4}{*}{\begin{tabular}[c]{@{}l@{}}Software \\ Maintenance\end{tabular}} & \multirow{4}{*}{66(36\%)} & \multirow{4}{*}{\begin{tabular}[c]{@{}l@{}}Code Comment \\ Quality Assurance\end{tabular}} & \multicolumn{2}{l|}{Code Comment Classification \& Quality Evaluation} & R{[}67, 84, 117{]} \\* \cmidrule(l){4-6} 
     &  &  & \multicolumn{2}{l|}{Code Comment Generation} & R{[}167{]} \\* \cmidrule(l){4-6} 
     &  &  & \multicolumn{2}{l|}{Obsolete Comment Identification} & R{[}94{]} \\* \cmidrule(l){4-6} 
     &  &  & \multicolumn{2}{l|}{Code Comment Updating} & R{[}101{]} \\* \midrule
    \multirow{10}{*}{\begin{tabular}[c]{@{}l@{}}General Task \\ Support\end{tabular}} & \multirow{10}{*}{25(14\%)} & \multirow{7}{*}{General Task} & \multicolumn{2}{l|}{Software/APP Classification} & R{[}4, 161{]} \\* \cmidrule(l){4-6} 
     &  &  & \multicolumn{2}{l|}{Tag Recommendation} & R{[}76{]} \\* \cmidrule(l){4-6} 
     &  &  & \multicolumn{2}{l|}{Sentiment Analysis/Detection} & R{[}6, 43, 85, 163{]} \\* \cmidrule(l){4-6} 
     &  &  & \multicolumn{1}{l|}{\multirow{4}{*}{\begin{tabular}[c]{@{}l@{}}Information \\ Retrieval\end{tabular}}} & Question Retrieval & R{[}15, 44{]} \\* \cmidrule(l){5-6} 
     &  &  & \multicolumn{1}{l|}{} & Answer Generation & R{[}115{]} \\* \cmidrule(l){5-6} 
     &  &  & \multicolumn{1}{l|}{} & Question Answering & R{[}166, 143, 169{]} \\* \cmidrule(l){5-6} 
     &  &  & \multicolumn{1}{l|}{} & Linkable Knowledge Prediction & R{[}97, 179{]} \\* \cmidrule(l){3-6} 
     &  & \multirow{3}{*}{Task Support} & \multicolumn{2}{l|}{SEthesaurus} & R{[}10{]} \\* \cmidrule(l){4-6} 
     &  &  & \multicolumn{2}{l|}{Bug Injection} & R{[}93, 180{]} \\* \cmidrule(l){4-6} 
     &  &  & \multicolumn{2}{l|}{Vector Representation} & \begin{tabular}[c]{@{}l@{}}R{[}12, 18, 63, 114, \\ 141, 147, 151{]}\end{tabular} \\* \midrule
    \multirow{10}{*}{\begin{tabular}[c]{@{}l@{}}Project \\ Management\end{tabular}} & \multirow{10}{*}{14(8\%)} & \multirow{3}{*}{Cost Management} & \multicolumn{2}{l|}{Software Development Effort Estimation} & R{[}56, 58{]} \\* \cmidrule(l){4-6} 
     &  &  & \multicolumn{2}{l|}{Change Impact Analysis} & R{[}33{]} \\* \cmidrule(l){4-6} 
     &  &  & \multicolumn{2}{l|}{Change-Prone Classes Prediction} & R{[}65{]} \\* \cmidrule(l){3-6} 
     &  & \multirow{2}{*}{\begin{tabular}[c]{@{}l@{}}Software Artifact \\ Link\end{tabular}} & \multicolumn{2}{l|}{Issue-Commit Links} & R{[}71{]} \\* \cmidrule(l){4-6} 
     &  &  & \multicolumn{2}{l|}{Traceability Link Recovery} & R{[}52, 113, 126, 164{]} \\* \cmidrule(l){3-6} 
     &  & \multirow{5}{*}{Others} & \multicolumn{2}{l|}{Authorship Analysis} & R{[}19{]} \\* \cmidrule(l){4-6} 
     &  &  & \multicolumn{2}{l|}{Task Recommendation} & R{[}69{]} \\* \cmidrule(l){4-6} 
     &  &  & \multicolumn{2}{l|}{Key Term Extraction} & R{[}148{]} \\* \cmidrule(l){4-6} 
     &  &  & \multicolumn{2}{l|}{Robust Malware Identification} & R{[}54{]} \\* \cmidrule(l){4-6} 
     &  &  & \multicolumn{2}{l|}{Commit Classification} & R{[}60{]} \\* \bottomrule[1.5pt]
    \end{longtable}
\end{landscape}

\subsubsection{SE Artifacts for WE Representation}

\begin{figure}[h]
    \centering
    \includegraphics[width=0.8\linewidth]{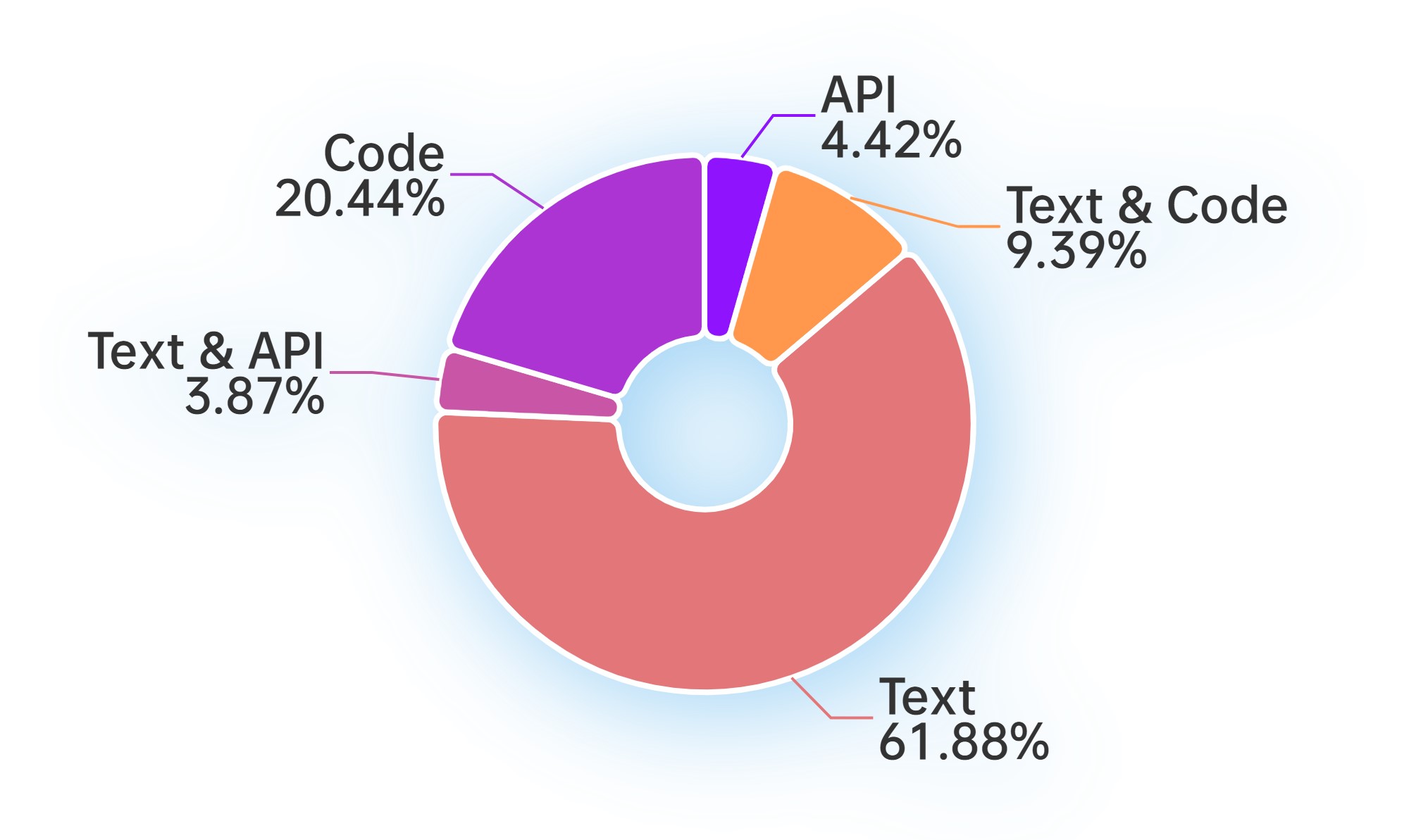}
    \caption{The distribution of software artifacts that use WE models for semantic representation in all 181 papers.}
    \label{fig:artifacts}
\end{figure}

The SE domain encompasses a wide range of software artifacts, including requirement documents, source code snippets, application programming interfaces (APIs), test cases, bug reports, log files, etc. Each of these artifacts contains rich semantic information, which are candidates which WE models can be applied to. In this part, we mainly aim to understand whether some kinds of SE artifacts are more likely to use WE models for semantic representation, and exactly what are they. To this, we read all the papers and marked the artifacts represented by WE models. Figure~\ref{fig:artifacts} shows the distributions of those artifacts.

From the figure, we can find that the distribution of software artifacts where WE models are applied to shows a strong connection with the distribution of software engineering (SE) tasks. The dominance of text-based artifacts (61.88\%) aligns with the high prevalence of WE technology usage in the software maintenance (36\%) and software development (24\%) areas, as both tasks heavily involve textual artifacts such as bug reports, requirement documents, and code documentation. Similarly, the significant use of WE in code-based artifacts (20.44\%) reflects its application in tasks like code analysis and bug localization. Artifacts that combine both text and code account for 9.39\%, while purely API-based ones represent 4.42\%. The smallest proportion, 3.87\%, comes from artifacts that combine text and APIs.

\begin{tcolorbox}
    [colback=gray!10,
    colframe=black!80,
    arc=2mm, auto outer arc,
    title={RQ2 - Summary},breakable,
    before upper={\parindent15pt\noindent},]
    
    Software maintenance (36\%) and software development (24\%) are the two areas within SE where WE models are most frequently applied. These two phases are also critical in the software lifecycle. Specifically, defect handling (42 studies) and code entity recommendation/generation (31 studies) are the two most dominant subareas. These tasks often involve substantial amounts of text (e.g., bug reports) and code, where capturing the underlying semantic information is crucial for effective problem-solving and automation. The distribution of software artifacts also mirrors this trend. Text-based artifacts dominate, comprising 61.88\% of the studied cases, followed by code-based artifacts at 20.44\%. This distribution reflects the fact that much of the work in these software engineering tasks involves processing textual descriptions of software issues or generating recommendations based on code, highlighting the potential of robust semantic understanding provided by WE models. The ability to effectively represent and interpret these artifacts can significantly enhance the accuracy and efficiency of various SE tasks.
\end{tcolorbox}

\subsection{RQ3. What WE models are generally adopted by SE tasks, and are they compared with other semantic representation models in the evaluation experiments?}

With this RQ, we hope to find out whether there exist any WE models that are commonly used by SE tasks, and further understand whether the authors of those studies performed relevant experiments to demonstrate the effectiveness of their adopted WE models by comparing with other semantic representation models. To answer RQ3, we carefully read each paper to identify the exact WE model adopted, and record any evaluation experiments of comparing the adopted WE model and other WE models or traditional semantic representation models like the VSM and topic model. If a paper adopted a WE model mainly inspired by findings from previous research and did not perform further comparison experiments, then we say it conducted a reference comparison. Following shows the detailed results.

\subsubsection{Distribution of various WE models used in SE tasks}

Figure~\ref{fig:popularity} demonstrate the frequency of use of various WE models in the SE field, reflecting their real-world applications and popularity. From the figure, we can find that:

\begin{figure}[h]
    \centering
    \includegraphics[width=0.8\linewidth]{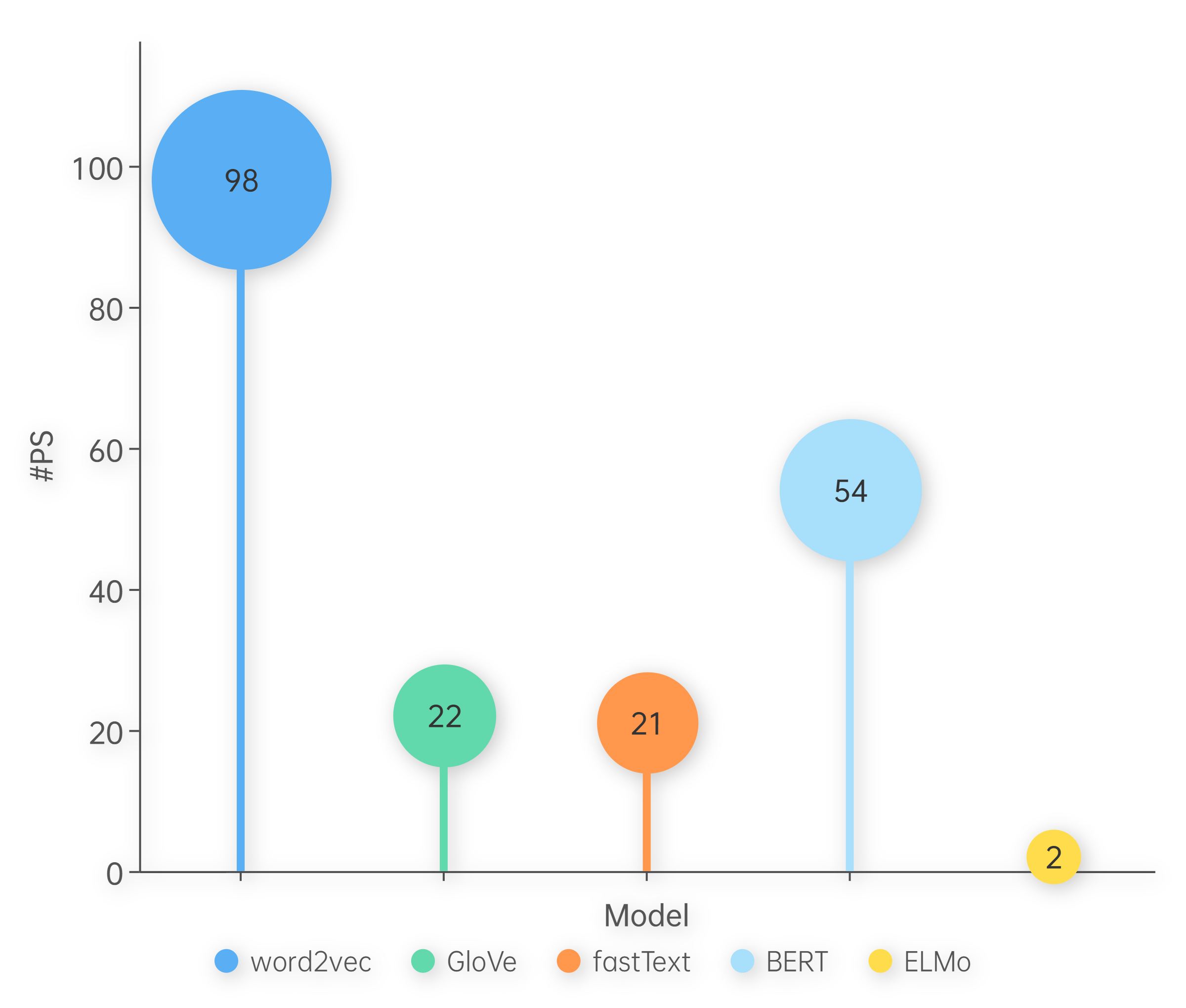}
    \caption{The distribution of different WE models in SE.}
    \label{fig:popularity}
\end{figure}

Word2Vec (98 related studies) is clearly the most commonly used word embedding model, likely due to its efficiency and adaptability to large-scale data. It effectively captures semantic relationships between words and performs well in SE tasks such as code documentation analysis and error detection.

Following Word2Vec, BERT (including its variants, such as CodeBERT) ranked second in usage, with 54 studies utilizing this bidirectional transformer model. BERT’s deep contextual understanding makes it highly suitable for SE tasks that require semantic comprehension, particularly in code understanding and natural language processing-related tasks.

FastText (21 studies), although less frequently used than Word2Vec and BERT, excels in handling out-of-vocabulary words and n-gram terms, giving it an advantage in SE scenarios where programming languages or terminologies are diverse. GloVe (22 studies) ranks slightly above FastText. As a model based on global word co-occurrence, GloVe performs well with large-scale text or code corpora but is less popular than Word2Vec and BERT.

In contrast, ELMo (2 studies), with its strong contextual capturing ability via bidirectional LSTMs, is used less frequently, possibly due to its higher computational resource requirements and longer training times compared to static word embedding models. Additionally, compared with the BERT model, which also requires considerable computational power , ELMo performs less effectively in handling complex sentences and long dependencies, making it less cost-effective overall.

\subsubsection{Comparison with other semantic representation models}
In this part, we mainly consider two kinds of comparisons, namely compared a WE model with traditional semantic representation models, and compared a WE model with other WE models. Result details are as follows.

\begin{itemize}
    \item[(1)] \textbf{Comparison with traditional models} 
\end{itemize}

Table~\ref{tab:traditional} shows the statistic results for studies that compared the adopted WE model with traditional semantic representation models. The studies did not perform such comparison are also counted. From the table, we can find that, while the use of WE in SE tasks is growing, 84\% of the studies (152 papers) did not compare them with traditional semantic representation methods. Only 9\% (17 studies) and 7\% (12 studies) conducted an experimental comparison or a reference comparison respectively. This low comparison rate suggests that the majority of research lacks direct performance evaluations between emerging word embedding techniques and traditional representation methods.

\renewcommand{\arraystretch}{1}
\begin{table}[h]
\centering
\caption{Comparison experiments between WE models and traditional word Semantic representation models}
\label{tab:traditional}
\begin{tabular}{@{}cccc@{}}
\toprule
Comparison Type                              & \#PS                      & Traditional Methods    & \#PS \\ \midrule
\multirow{10}{*}{Experimentation Comparison} & \multirow{10}{*}{17(9\%)} & TF-IDF                 & 13   \\
                                             &                           & one-hot                & 2    \\
                                             &                           & LDA                    & 2    \\
                                             &                           & BM25                   & 3    \\
                                             &                           & BOW                    & 1    \\
                                             &                           & topical word embedding & 1    \\
                                             &                           & WMD                    & 1    \\
                                             &                           & Jaccard                & 1    \\
                                             &                           & Edit                   & 1    \\
                                             &                           & Bi-Gram                & 1    \\
\multirow{6}{*}{Reference Comparison}        & \multirow{6}{*}{12(7\%)}  & one-hot                & 6    \\
                                             &                           & WordNet                & 2    \\
                                             &                           & TF-IDF                 & 1    \\
                                             &                           & LSA                    & 1    \\
                                             &                           & n-Gram                 & 1    \\
                                             &                           & BOW                    & 1    \\
No Comparison                                & 152(84\%)                 & /                      & /    \\ \bottomrule
\end{tabular}
\end{table}

The above results indicate that there lacks convincing experimental evidence that supports the decision of existing studies that use WE models instead of traditional and relatively simpler semantic representation models to represent the semantics of SE tasks. Considering that the most suitable semantic representation models of different tasks and scenarios may also be different. It would be appreciated if systematic evaluations between WE models and traditional models over various SE tasks are performed. Such comparisons will provide researchers with a more comprehensive theoretical foundation and empirical support, promoting more rational and informed decisions when selecting semantic representation models for their SE tasks at hand.

\begin{itemize}
    \item[(2)] \textbf{Comparison with other WE models} 
\end{itemize}

After we reviewed all 181 papers, we found that 119 studies (66\%) did not conduct any direct comparisons between word embedding models. Among the remaining 62 papers, 22 papers referenced related literature to explain their choice of WE models, and only 40 studies (22\%) performed experimental comparisons between WE models to demonstrate the optimal choice of certain WE models. The detailed comparison between different WE models are shown in Table~\ref{tab:vs_WE}.

\begin{table}[h]
\centering
\caption{Comparison experiments between WE models}
\label{tab:vs_WE}
\begin{tabular}{ccc}
\toprule
Adopted Model                        & PS                  & Compared WE models                                   \\ \midrule
\multirow{5}{*}{Word2Vec}            & \multirow{5}{*}{11} & Doc2vec                                              \\
                                     &                     & BERTbase                                             \\
                                     &                     & FastText                                             \\
                                     &                     & FastText, Doc2vec                                    \\
                                     &                     & BERTbase, ELMo                                       \\ \hline
\multirow{4}{*}{BERT \& variants}    & \multirow{4}{*}{8}  & Word2Vec                                             \\
                                     &                     & GloVe                                                \\
                                     &                     & BERTbase                                             \\
                                     &                     & Transformer                                          \\ \hline
\multirow{3}{*}{GloVe}               & \multirow{3}{*}{5}  & Word2Vec                                             \\
                                     &                     & Word2Vec, FastText                                   \\
                                     &                     & Word2Vec, sentiment specific word   embedding (SSWE) \\ \hline
\multirow{4}{*}{FastText}            & \multirow{4}{*}{4}  & Word2Vec                                             \\
                                     &                     & Word2Vec, GloVe                                      \\
                                     &                     & BERTbase, Sent2Vec                                   \\
                                     &                     & GloVe, Doc2vec                                       \\ \hline
\multirow{2}{*}{ELMo}                & \multirow{2}{*}{2}  & CBOW                                                 \\
                                     &                     & Word2Vec, GloVe, FastText                            \\ \hline
CodeBERT                             & 1                   & CodeBERT(without finetuning)                         \\ \hline
\textit{skip-gram,   FastText}                & \textit{1} & \textit{skip-gram,   FastText}                  \\ \hline
\textit{Word2Vec,   GloVe}        & \textit{1}          & \textit{Word2Vec,   GloVe}                           \\ \hline
\textit{Word2Vec,   GloVe}        & \textit{1}          & \textit{Word2Vec,   GloVe}                           \\ \hline
\textit{Word2Vec,   GloVe, FastText}          & \textit{1} & \textit{Word2Vec,   GloVe, FastText}            \\ \hline
\textit{Word2Vec,   GloVe, FastText,BERTbase} & \textit{1} & \textit{Word2Vec,   GloVe, FastText,  BERTbase} \\ \hline
\textit{Word2Vec,   BERT\&variants}           & \textit{4} & \textit{Word2Vec,   BERT\&variants}                  \\ \bottomrule        
\end{tabular}
\end{table}

The results in Table~\ref{tab:vs_WE} reveal that a variety of WE models have been compared across different studies, with Word2Vec and BERT\&variants being the most frequently tested , appearing in 11 and 8 studies, respectively. Word2Vec was compared against models such as Doc2Vec, FastText, and BERTbase, highlighting its versatility and competitive performance in different SE tasks. Similarly, BERT and its variants, were commonly compared with models like GloVe, Transformer, and other WE models, showcasing the growing interest in contextual embeddings for SE applications. GloVe appeared in 5 studies, often compared with Word2Vec, FastText, and specialized embeddings like Sentiment-Specific Word Embedding (SSWE). FastText was evaluated in 4 studies, where it was compared with models like Word2Vec, GloVe, and contextual embeddings like BERTbase and Sent2Vec, reflecting its strength in handling out-of-vocabulary words and morphological variations in software text. Additionally, models such as ELMo and CodeBERT were less frequently compared, appearing in 2 and 1 studies, respectively, but their inclusion suggests a focus on more advanced contextual models and domain-specific embeddings for tasks like source code understanding. These comparisons reflect an evolving focus on understanding the performance differences between classical and contextual embedding models, particularly in SE-specific applications. While Word2Vec and BERT remain popular choices, a broader exploration of specialized models is taking place, indicating a shift toward more targeted and fine-tuned approaches in software engineering tasks. 

Additionally, the italicized portions in the Table~\ref{tab:vs_WE} indicate that, although these studies (a total of 9) conducted comparative experiments between different WE models, they did not ultimately select any one model as their semantic representation method for SE tasks. This suggests that, while a variety of word embedding models were evaluated, the final choice of model for representing SE data was left open, possibly due to the models' varying performances across different tasks, or due to a preference for using multiple models for comprehensive analysis. This further highlights the complexity of selecting a single, optimal word embedding model for SE tasks, especially when balancing the trade-offs between performance and task-specific requirements.

\begin{tcolorbox}
    [colback=gray!10,
    colframe=black!80,
    arc=2mm, auto outer arc,
    title={RQ3 - Summary},breakable,
    before upper={\parindent15pt\noindent},]
    
    Word2Vec is the most widely used word embedding model in software engineering, followed by BERT, reflecting their adaptability and effectiveness in handling large-scale data and complex semantic tasks. However, exploration of newer, task-specific models remains limited compared to these more mature models. Despite the widespread adoption of word embeddings, 84\% of studies did not compare them with traditional representation methods such as TF-IDF or one-hot encoding. Without consistent comparisons, it is difficult to fully assess the added value of word embeddings over simpler methods in SE applications. Similarly, only 22\% of studies conducted experimental comparisons between different WE models, indicating a gap in systematically exploring which model performs best for specific SE tasks.
\end{tcolorbox}

\subsection{RQ4. What is the general way to obtain WE vectors in SE tasks? By using the general pre-trained WE models or training a domain-specific one?}

Generally speaking, there are two ways to obtain word embeddings. One is to directly use a general pre-trained WE model, which has already been trained on large external datasets (like Wikipedia or Google News); the other one is to use domain-specific datasets to train a WE model from scratch or fine-tune a general pre-trained model to generate the embedding vectors. This RQ could help us understand the current practice of obtaining WE vectors and identify potential improvements. To answer RQ4, we employed the following approach. If the corpus used for training the WE model was general web contents like Wikipedia or Google News, then we say the specific study used a generic pre-trained WE model. Conversely, if the corpus was specific to the software engineering (SE) domain (e.g., Stack Overflow), then we say a domain-specific, i.e., SE-specific, WE model is used.

\begin{figure}[h]
    \centering
    \includegraphics[width=0.8\linewidth]{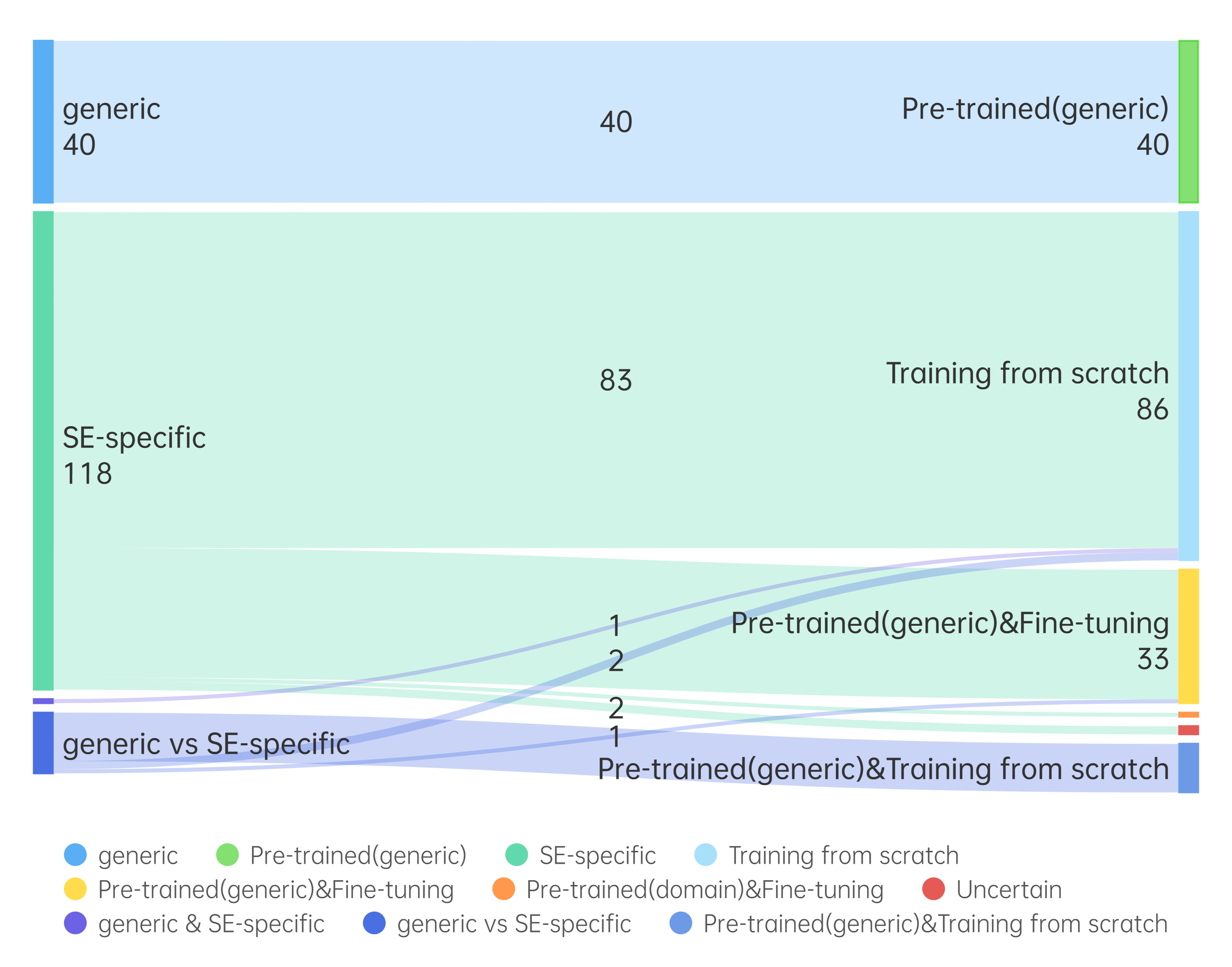}
    \caption{The type and training methods of WE in SE.}
    \label{fig:training}
\end{figure}

Most papers would explicitly state the corpus used for training WE models and its source. However, in some studies, the word embedding model was described without clearly specifying the corpus. In such cases, we followed the links provided by the authors to the word embedding model to determine whether it was generic or domain-specific. After checking all papers, the following WE generation strategies are used in different studies. Figure~\ref{fig:training} shows the detailed statistics.

From Figure~\ref{fig:training}, we can find that, except for 9 papers that did not clearly state the generation details of WE vectors, the majority (118 papers) utilized SE-specific embeddings trained on domain-specific corpora. Among these, 83 studies trained the models from scratch on SE-related datasets, while 33 studies applied fine-tuning methods, adapting pre-trained generic models to better suit SE-specific tasks. This shows a significant preference for customizing embeddings to the SE domain, likely due to the specialized vocabulary and unique linguistic patterns found in software-related texts. In contrast, 40 studies used generic word embeddings trained on large, non-SE-specific corpora (e.g., Wikipedia). These models were pre-trained and directly applied without further customization. Additionally, 16 studies trained word embeddings using both general corpora and SE-specific corpora, among which 15 studies conducted experiments to compare their performance. Eleven studies reported that SE-specific embeddings outperformed general embeddings, while the other four indicated no significant difference in their performance. Another study, due to its task specificity (i.e. SEthesaurus), mixed a general corpus and a specialized corpus for word embedding training.

\begin{tcolorbox}
    [colback=gray!10,
    colframe=black!80,
    arc=2mm, auto outer arc,
    title={RQ4 - Summary},breakable,
    before upper={\parindent15pt\noindent},]
    
    Most SE tasks chose to use domain-specific data to train a WE model or fine-tune a pre-trained general one to obtain the WE vectors of their SE artifacts. Yet, few studies have tried to conduct comparative experiments to check whether the domain-specific WE performs better than the general pre-trained one. it would be interesting and very valuable to conduct further investigation into when and where domain-specific embeddings provide a tangible advantage over general pre-trained ones, or vice versa.
\end{tcolorbox}

\section{Implications}
\label{discuss_sect}
Our work systematically reviews the application of WE models in the SE domain by answering four research questions related to the prevalence of WE models across various SE tasks and software artifacts, the selection of pre-trained or domain-specific models, etc. 
Based on the results, we have summarized some actionable research opportunities as follows. Progress in these research directions can further advance the development of software technology and semantic representation techniques.

\begin{itemize}
    \item \textbf{Expanding the Application of WE Models Beyond Unstructured Textual Artifacts and in Less-Explored SE Areas.}
    WE models have gained significant recognition in major academic journals and conferences (in RQ1) and are applied in various SE tasks, primarily on software development and maintenance tasks, and focused mainly on the semantic representation of unstructured textual artifacts (in RQ2).
    To harness the full potential of WE models, future research should aim to expand their use into less-explored tasks/areas like requirements engineering, testing, and design, thereby increasing their applicability and driving innovation across the broader landscape of software engineering.
    Furthermore, it is also essential to extend the application of WE models beyond non-structured software artifacts to other forms, including structured data, graphical control/data flow, formal requirements specification, etc. This would enhance their adaptability to a wider range of tasks within software engineering and enable more comprehensive support for automation, analysis, and optimization across the entire development process.

    \item \textbf{Enhancing Comparative Analysis of WE Models and Traditional Semantic Retrieval Techniques.} After thoroughly reviewing all 181 primary studies, we found a notable scarcity of comparative analyses between WE models and traditional semantic retrieval methods such as TF-IDF and one-hot encoding of the vector space model (in RQ3). Only a small percentage of these studies have conducted systematic comparisons, which hinders our ability to fully evaluate the advantages and disadvantages of employing WE models in place of simpler, well-established techniques for various SE tasks.
    Moreover, there is a significant lack of comparative studies among different WE models themselves, resulting in an incomplete understanding of which models are most effective for addressing specific challenges within the SE domain. This gap highlights the need for future research to prioritize comprehensive evaluations that compare WE models against traditional methods as well as among themselves across a diverse range of SE scenarios. Such investigations will not only clarify the performance dynamics of these models but also guide practitioners in selecting the most suitable semantic retrieval models for their specific needs.

    \item \textbf{Advancing WE Models for Enhanced Performance in SE Tasks.} The predominance of Word2Vec, followed by BERT and its variants, indicates a clear preference for models that excel in capturing semantic relationships and contextual understanding (in RQ3). This trend suggests several promising research directions aimed at enhancing the efficacy of WEs in SE tasks. One could be the integration and hybridization of existing models. For example, by leveraging the strengths of Word2Vec's efficiency and BERT's deep contextual understanding, researchers can develop novel embedding models that maximize performance across a range of SE applications, such as code documentation analysis and error detection. This could involve creating composite models that combine the rapid processing capabilities of traditional methods with the advanced semantic comprehension of newer transformer-based architectures. Another one involves addressing the challenges posed by diverse programming languages and terminologies. Models like FastText and GloVe, though less frequently used, offer unique advantages in handling out-of-vocabulary words and large-scale text corpora. Research efforts could focus on refining these models or integrating them with more popular frameworks to create robust solutions tailored to the complex linguistic structures found in SE contexts. 

    
    \item \textbf{Comparing Domain-Specific and General Word Embeddings in the SE Domain.} The clear preference for SE-specific embeddings over general embeddings (in RQ4), such as those trained on domain-specific corpora like Stack Overflow, underscores a growing awareness of the significance of domain-specific knowledge in addressing SE tasks. Despite this shift towards specialization, a notable gap remains in the systematic research that directly compares SE-specific embeddings with their general counterparts. Some studies suggest that the differences between SE-specific and general embeddings may not always be significant for certain tasks, opening up avenues for deeper investigation. It is crucial to discern the contexts and conditions under which domain-specific embeddings provide meaningful advantages over general embeddings. Such explorations could yield valuable insights, ultimately contributing to the development of clearer guidelines for practitioners and researchers in selecting between general pre-trained WE models and domain-specific WE models tuned/trained with SE data. By tailoring model choices to the unique demands of various tasks, we can enhance the effectiveness and efficiency of solutions in the SE domain.
\end{itemize}

\section{Threats to Validity}
\label{threat_sect}
\subsection{Internal Validity. }
A potential threat lies in the process of selecting relevant studies. While we did not thoroughly read the entire content of every paper, we followed a systematic and structured approach. Specifically, we assessed papers by examining the title, abstract, and introduction. In cases where these sections did not provide sufficient clarity, we resorted to reviewing the full paper. While this method may introduce some bias by potentially missing details buried deeper in the text, it is a commonly accepted practice in literature reviews. This approach allows for efficient filtering while maintaining a high level of rigor, ensuring that we included relevant studies that aligned with our research focus.
Another internal validity concern arises from the classification of SE tasks. Since the categorization of tasks is subject to human judgment, there is an inherent risk of subjectivity in the process. To mitigate this, we followed established frameworks and task definitions from prior literature whenever possible. We also ensured that our classification was consistent across studies by having multiple reviewers discuss and agree on the task categories. By employing this collaborative approach, we reduced the potential for bias and improved the reliability of the classifications.

\subsection{External Validity. }
One threat is the limitation of our literature sources to CCF A and B-ranked conferences and journals specifically within the Software Engineering (SE) domain. The China Computer Federation (CCF) provides a comprehensive directory that classifies high-quality scientific journals and conferences based on their impact and reputation in various fields, including SE. These CCF A and B rankings are recognized for identifying the most prestigious venues, including internationally renowned conferences such as ICSE, FSE, and journals like IEEE Transactions on Software Engineering. By focusing on these high-ranking venues, we aimed to ensure that the studies included in our review were of the highest academic and research standards, reflecting significant contributions to the field. However, this focus may inadvertently exclude valuable insights from lower-tier venues. Future reviews could consider incorporating studies from a wider range of sources to provide a more comprehensive view of the SE landscape.
Besides, this review focuses solely on peer-reviewed academic publications, potentially excluding relevant insights from non-academic or "gray" literature, such as technical blogs, white papers, and industry reports. This exclusion might overlook the most current practical innovations and trends in the SE industry. Nonetheless, peer-reviewed literature offers a more rigorous validation of findings, ensuring that the included studies meet a certain standard of quality. To address this limitation, future reviews could consider a more systematic inclusion of high-quality gray literature to capture cutting-edge practices in the field.

\section{Related Work}
\label{related_sect}
Related to our work of studying the use of WE in SE domain, some studies have tried to provide a comprehensive study on the applications of other advanced techniques in the SE domain.  For example, in\cite{19}, the application of Machine learning (ML) throughout the SE lifecycle has been widely explored, particularly in the areas of software quality and testing, but challenges remain in fields such as human-computer interaction. Yang et al.\cite{20} summarized the effectiveness and challenges of Deep Learning (DL) in various SE tasks. The application of DL in software testing and maintenance, such as defect prediction and code analysis, has shown significant potential\cite{21}. Wang et al.\cite{22} also discussed the impact of ML and DL on SE, particularly the issues of complexity and reproducibility. Large language models (LLMs) have also demonstrated potential in optimizing SE processes and outcomes\cite{23}. The above studies together with our study, could provide better and more broader support on cross-discipline technique adoption in facilitating the development of SE techniques.

There also exist some studies from the NLP area that mainly systematically evaluated various kinds of WE models in certain NLP tasks. For example, In\cite{24}, the authors found that WE techniques could significantly improve the performance of text classification techniques, and largely surpassed traditional bag-of-words models. Some researchers tried to divide existing WE models into traditional word embedding, static word embedding, and contextualized word embedding, and emphasized the significant performance advantages of BERT, in tasks like sentiment classification, text classification and next sentence prediction\cite{25}. In a systematic review of\cite{26}, neural network-based WE methods are found to outperform matrix factorization techniques (i.e., variants of word2vec). Some researchers explored the theoretical foundations and development trajectory of WE, and analyzed the advantages of using WE models in semantic representation\cite{27,28}. By optimizing training methods and corpus selection, WE techniques could significantly improve classification accuracy in sentiment analysis\cite{29,30}. Cross-lingual WE\cite{31} are also found to be able to improve the accuracy of semantic reasoning in multilingual environments.  Incitti et al.\cite{32} conduct a performance evaluation of text embedding models, with a focus on embedding such as textual sentence or paragraph that go beyond words.

The above representative studies either focus on completely different technology adoption in the SE domain, or are mostly limited to the comprehensive study of WE use in the NLP field, with little attention being devoted to studying the current practice of using WE in the broad SE domain. To fill this research gap, this paper aims to provide a more comprehensive perspective through a systematic review of the applications of WE techniques in the field of SE. Unlike existing studies, we do not restrict our focus to specific SE tasks but instead broadly collect and discuss research that employs word embeddings WE as a semantic representation method across the field of SE. This will help researchers better understand and master techniques for the semantic representation and processing of SE artifacts, such as code and requirements documents, providing strong support for the further development of tasks related to semantic representation in SE.

\section{Conclusion}
\label{conclude_sect}
In this study, we conducted a comprehensive study to understand the adoption of WE models in the software engineering domain, by systematically analyzed 156 primary studies. Related to this, we first checked the publication trend of these studies. Then, we explored the prevalence of WE model usage across various SE tasks and software artifacts. After that, we studied the dominant WE models used in SE tasks, analyzed the rationale behind the choice of certain WE models by checking the existence of comparison experiments with other semantic representation models. Last, we checked their training strategies to understand the leveraging of domain specific knowledge in model training. Based on the analysis of the results, we obtained a general view of the adoption practice of WE in SE, and also identified several key challenges and future research directions.

\bibliographystyle{unsrt} 
 

\end{document}